\documentclass[prb,twocolumn,superscriptaddress, showpacs,floatfix]{revtex4}

\usepackage[dvips]{graphicx}
\usepackage{amssymb, amsmath, bbm}

\def\be{\begin{eqnarray}}
\def\ee{\end{eqnarray}}

\pacs{MPI-PKS-Dresden, preprint ID:  242530.0}

\begin{document}
\title{Magnetic phase diagram of a frustrated  ferrimagnetic ladder:
Relation to the one-dimensional boson Hubbard model}
\author{N. B.  Ivanov}
\affiliation{ Max-Planck Institut f\"ur Physik Komplexer Systeme,
N\"othnitzer Strasse 38, D-39016 Dresden, Germany}
\altaffiliation[Permanent address:]{ Institute of Solid State Physics, 
Bulgarian Academy of
Sciences, Tzarigradsko chaussee 72, 1784 Sofia, Bulgaria}
\author{J. Richter}
\affiliation{Institut f\"ur Theoretische Physik, Universit\"at Magdeburg,
PF 4120, D-39016 Magdeburg, Germany}
\date{\today}
\begin{abstract}
We study the magnetic  phase diagram of   two coupled  
mixed-spin $(1,\frac{1}{2})$  Heisenberg chains as a function of 
the frustration parameter related to diagonal exchange couplings. 
The analysis is performed by using  spin-wave series  and
exact numerical diagonalization techniques.
The obtained phase diagram --  containing the Luttinger liquid  
phase, the  plateau phase with a  magnetization per 
rung $M=1/2$, and the  fully polarized phase -- 
is closely related to the generic 
$(J/U,\mu/U)$ phase diagram of the  one-dimensional
boson Hubbard model.
\end{abstract}
\maketitle

Some special  classes  of quantum  spin ladders  composed of  
different types of  spins 
and/or   regular mixtures  of ferromagnetic and antiferromagnetic 
exchange bonds  exhibit a number of novel quantum spin phases  
and unusual thermodynamic properties.\cite{1,2} 
Moreover, such systems are appropriate for
studying magnetic quantum critical points in one spatial 
dimension (1D), since many of them sustain 
magnetically ordered ground states.  
Possible realizations of such ladder structures can be based, for
instance, on the synthesized  quasi-1D bimetallic  
molecular magnets.\cite{kahn,hagiwara} 

In a recent paper  we  considered the role of   
geometric  frustration in a system of two coupled  
mixed-spin  $(1,\frac{1}{2})$ Heisenberg chains.\cite{ivanov1} 
It has been argued that a relatively moderate strength of the
magnetic frustration  produced by the nearest-neighbor diagonal bonds 
was  able to  destroy the ferrimagnetic phase   
and to stabilize the  Luttinger liquid  phase. 
The quantum phase transition between these two phases is smooth
and is  realized  through an intermediate  canted spin  phase.
This latter phase is characterized
by a  net ferromagnetic moment per rung, $0< M <0.5$, 
 and power-law transverse spin-spin correlations.
In this Brief Report we show that the  magnetic  phase diagram 
of the discussed  model as a function of the 
frustration parameter  is closely related to the $(J/U,\mu/U)$
phase diagram of the boson Hubbard model 
in periodic 1D potentials,\cite{fisher}
where $\mu$, $J$, and $U$ are, respectively,
the chemical potential, the
hopping strength constant, and 
the on-site repulsion potential.

The relevant spin Hamiltonian reads
\begin{multline}\label{h}
{\cal H}=\sum_{n=1}^L\left[ 
J_1\left({{\bf s}_{1}}_n\cdot {{\bf s}_{2}}_{n+1}
+{{\bf s}_{2}}_n\cdot {{\bf s}_{1}}_{n+1}\right)
+J_{\perp}{{\bf s}_{1}}_n\cdot {{\bf s}_{2}}_{n}\right.\\
+J_2 
\left.\left({{\bf s}_{1}}_{n}\cdot {{\bf s}_{1}}_{n+1}+ 
{{\bf s}_{2}}_{n}\cdot {{\bf s}_{2}}_{n+1}\right) 
-g\mu_BH({s_1}_n^z+{s_2}_n^z)\right]
\, ,
\end{multline}
where  $L$, $g$, $\mu_B$, and $H$  are, 
respectively, the number of rungs,
the gyromagnetic ratio, the Bohr magneton, and the external
magnetic field applied in the $z$ direction. The spin operators
${{\bf s}_{1}}_n$ and ${{\bf s}_{2}}_n$ (characterized 
by the  quantum spin numbers $s_1>s_2$ and the rung index $n$) 
are alternatively distributed on the ladder sites.
The parameters $J_1$, $J_{\perp}$, and $J_2$ are positive  coupling
constants of the nearest-neighbor horizontal, vertical, and 
diagonal exchange bonds, respectively (see Fig. 1 in Ref.
\onlinecite{ivanov1}). We set the energy 
and length scales by $J_1\equiv 1$ and $a_0\equiv 1$,
where $a_0$ is  the lattice spacing. 
\begin{figure}[hbt]
\samepage
\begin{center}
\includegraphics[width=8cm]{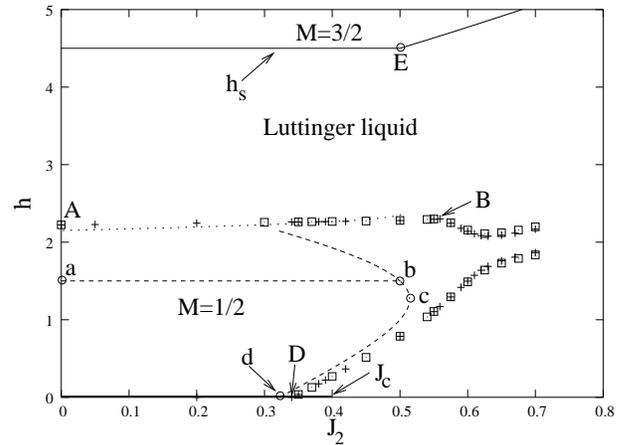}
\caption{\label{fig1}
Phase diagram of the model (\ref{h}) in the $(J_2,h)$ plane.
The phase boundary $h_s=h_s(J_2)$ of the fully polarized phase 
($M=3/2$) is exact. The dashed curve $abcd$
marks the area occupied by the $M=1/2$ plateau phase, 
as obtained in a linear 
spin-wave approximation. The boundary position of this
phase,  as obtained from  exact numerical diagonalization of
periodic clusters, is traced  by squares
($L=10$) and crosses ($L=12$).   
The dotted curve  shows the phase boundary where 
$\Delta^{(-)}_{0}(h)=0$ [see Eq. (\ref{omega3})], 
as obtained from the first-order spin-wave theory.
The dashed curve starting from the point $b$ and 
going upwards left as well as the piece $bc$
represents in a linear spin-wave  approximation the 
boundary where $\Delta^{(-)}_{\pi}(h)=0$.
The canted spin phase exists for  $h=0$
in the interval $0.342<J_2<J_c$, where $J_c=0.399$
is the phase transition point from the  canted phase
to the Luttinger liquid phase.
 The rest of the area is occupied by the 
Luttinger liquid phase. 
}
\end{center}
\end{figure}
For simplicity, in the following  we consider the case $J_{\perp}=1$
and set $h=g\mu_BH$.

The magnetic phase diagram of the model (\ref{h}) in the 
($J_2,h$) plane is presented in Fig. \ref{fig1}.   
The diagram contains the  Luttinger liquid  phase, the
plateau magnetic phase with a quantized magnetic moment per 
rung $M=1/2$ and the fully polarized phase with $M=3/2$. 
In addition, there is a canted magnetic phase which exists
only at zero magnetic field (in the interval 
$0.342<J_2<J_c=0.399$)  and for $h>0$  merges 
into the Luttinger liquid  phase. Some properties of the latter phase
at $h=0$ have already been studied in Ref. \onlinecite{ivanov1},
by using  a conformal-field-theory analysis of  
the numerical exact-diagonalization (ED) data 
for the ground-state energy and the energies of 
some low-lying excited states in finite periodic systems. 
As is well known,  the  external magnetic field varies the parameters of 
the Luttinger liquid phase without changing its basic properties up to
a saturation field ($h_s$), where the system becomes fully 
polarized.\cite{konik} 

In this connection, let us first consider the  critical field 
$h_s$ defining the phase boundary of the fully polarized $M=3/2$ phase. 
The boundary $h_s=h_s(J_2)$ can  be obtained by  examining 
the instabilities of the one-magnon excitations above the ferromagnetic
vacuum. We assume, as generally accepted, 
that the multimagnon excitations become unstable  
at a weaker magnetic field as compared to the  one-magnon excitations. 
As a matter of fact, the latter  assumption is confirmed by our 
ED data, which exactly reproduce the analytical result
for $h_s$ presented below.
A  straightforward calculation results in 
the following exact expression for the  one-magnon
dispersion relations in the fully polarized phase
\begin{multline}\label{omega1}
\omega^{(1,2)}(k)=h-\frac{9}{4}-3J_2\sin^2\left(\frac{k}{2}\right)\\
\pm \frac{1}{2}\sqrt{\left[ \frac{3}{2}-2J_2\sin^2
\left(\frac{k}{2}\right)\right]^2
+2\left[ 3-4\sin^2\left(\frac{k}{2}\right)\right]^2}
\, ,
\end{multline}
where the   wave vector $k$  runs in  the lattice Brillouin zone
$-\pi \leq k < \pi$. 
Equation (\ref{omega1}) describes two folding ferromagnetic (FM)
branches of  one-magnon excitations.
 The  phase  boundary $h_s=h_s(J_2)$  is defined by
the expressions
\begin{eqnarray}\label{hs}
h_s&=&\frac{9}{2} \hspace{0.2cm} (0\leq J_2\leq 0.5)
\nonumber\\
   &=&\frac{9}{4}+3J_2
+\frac{1}{2}\left[ \left( \frac{3}{2}-2J_2\right)^2+2
\right]^{1/2} \hspace{0.2cm}
(J_2 \geq  0.5),  
\nonumber
\end{eqnarray}
where the  first and second lines  correspond, respectively,
to instabilities of $\omega^{(2)}(k)$
at the wave vectors $k=0$ and $\pi$. For  $h<h_s(J_2)$, the modes 
around $k=0$ and $\pi$ become unstable and destroy  the fully polarized state. 
As discussed in the following text, the point $E$ in Fig. \ref{fig1}
is a crossover point at which
both  minima in $\omega^{(2)}(k)$  reach the ground state.   

Let us now address  the $M=1/2$ plateau phase.
The presence of quantum fluctuations 
in the ferrimagnetic state  does not allow 
exact considerations, so that we shall 
rely on  a qualitative analysis based on the 
spin-wave theory (SWT) supplemented by numerical 
ED data for finite periodic
systems containing up to $L=12$ rungs. 
In a linear SWT approximation, 
the magnon  excitations above the
classical ferrimagnetic ground state are described
by the following dispersion relations:  
\begin{multline}\label{omega2}
\omega^{(\pm)}(k)=\pm h\mp\left[\frac{3}{4}
+J_2\sin^2\left(\frac{k}{2}\right)\right]\\
+\frac{1}{2}\sqrt{9\left[ \frac{3}{2}-2J_2
\sin^2\left(\frac{k}{2}\right)\right]^2
-2\left[ 3-4\sin^2\left(\frac{k}{2}\right)\right]^2}
\, .
\end{multline}
In the last equation, $\omega^{(+)}(k)$ is a FM branch,  
whereas $\omega^{(-)}(k)$ describes  antiferromagnetic (AFM) 
one-magnon excitations. 
In the  whole region occupied by the $M=1/2$ phase, the
dispersion relations (\ref{omega2}) exhibit
minima at the center ($k=0$) as well as at the boundary ($k=\pi$) 
of the Brillouin zone. Close to  these wave vectors 
the magnon energies take the generic form
\be\label{omega3}
\omega^{(\pm)}_{k_0}(k)=\Delta^{(\pm)}_{k_0}(h)
+\frac{(k-k_0)^2}{2m^{(\pm)}_{k_0}}
\,  (k_0=0,\pi)\, .
\ee
Here  $\Delta^{(\pm)}_{k_0}(h)$ and $m^{(\pm)}_{k_0}$ are, respectively,
the magnon gaps and effective masses. Since $\omega^{(+)}_{0}(k)$ is 
a  Goldstone mode, we have $\Delta^{(+)}_{0}(0)=0$.

The  magnon modes in  Eq. (\ref{omega3}) 
become gapless  on  different parts  of the   phase boundary 
marked by the points $\it{a}$, $\it{b}$, $\it{c}$, and $\it{d}$;
$\omega^{(-)}_{0}(k)$, $\omega^{(-)}_{\pi}(k)$,
and $\omega^{(+)}_{\pi}(k)$ are gapless, respectively,
on the pieces $\it{ab}$, $\it{bc}$, and $\it{cd}$.  
At the special  point $\it{b}$ (analogous to the point $E$), 
both  AFM  modes  $\omega^{(-)}_0(k)$ and  $\omega^{(-)}_{\pi}(k)$ 
are gapless. Similar  special points 
have   been found  in  other 1D spin models with 
folding excitation spectra.\cite{okunishi}  In the case when the
related  magnon masses in Eq.~(\ref{omega3}) diverge, such points 
produce cusps in the magnetization curves. Otherwise, 
as in our case, these are crossover points marking the boundary between
regions with different low-lying excitations which belong 
to a single excitation branch. On the other
hand, at the special point   $\it{c}$ with coordinates
$(J_2^*,h^*)=(0.75-\sqrt{2}/6,1.5-\sqrt{2}/6)$, 
two different types of modes --
the FM mode  $\omega^{(+)}_{\pi}(k)$ and the AFM 
mode $\omega^{(-)}_{\pi}(k)$ --  reach the ground state.
The point $\it{c}$ belongs to the special line $h=0.75+J_2$ on which   
the energies of both  magnon modes coincide:   
$\omega^{(+)}_{\pi}(k)=\omega^{(-)}_{\pi}(k)$. 
Using the picture of  "particle-hole"  excitations
defined by  $\omega_{ph}(k)=\omega^{(+)}(k)+\omega^{(-)}(k)$,
we find that close enough to ($J_2^*,h^*$) the dispersion
 relation  $\omega_{ph}(k)$ takes the  relativistic form
\be\label{eph}
\omega_{ph}(k)=\sqrt{\Delta_{ph}^2+v_{ph}^2\left( k-\pi\right)^2}\, .
\ee
While the velocity $v_{ph}$ remains finite at ($J_2^*,h^*$),
the excitation gap vanishes as $\Delta_{ph}\propto 
\left(J_2^*-J_2\right)^{1/2}$. Clearly, the discussed structure
of the low-lying excitations in a vicinity of ($J_2^*,h^*$) 
remembers the known structure of low-lying 
excitations  close to the  commensurate  
Mott-insulator-superfluid  transition.\cite{fisher}

In fact, the outlined  peculiarities of the 
$(J_2,h)$ phase diagram --
basically deduced from  the linear SWT approximation --        
share  a number of common features with  the $(J/U,\mu/U)$  
phase diagram of the   boson Hubbard model
on 1D periodic lattices. Before  discussing this issue, 
let us  present  our quantitative estimates 
for the boundary  positions  based on  higher-order spin-wave 
series and numerical ED data. 
Unfortunately, already the first-order corrections in the 
SWT series  exhibit divergences of the form $\delta^{-1/2}$, 
$\delta$ being the distance in the parametric $(J_2,h)$ space
from  the point ($J_2^*,h^*$).  Thus, reliable  quantitative 
results can be achieved only for large enough $\delta$. 
As an example, we show in Fig. \ref{fig1} 
the first-order SWT result for the  critical field $h_c$ defined
through the relation $\Delta^{(-)}_0(h_c)=0$. 
In spite of the relatively large  
correction of the boundary position, we observe
a good  agreement  with the ED data.
In fact, one may expect further improvement of 
the theoretical  result in a second-order SWT  
approximation.\cite{sen}  
Relatively large  finite-size effects in the
ED data  can be indicated only  in the vicinity of the 
tip of the region occupied by the $M=1/2$ phase.  
It is remarkable that the  peculiarities
of the phase  boundary -- found in the framework
of the  linear SWT approximation --  
are  reproduced by the ED data. 
In particular, the cusp at  point $B$ -- signaling a
change in the type of low-lying excitations --
can easily be  indicated by the ED data. This is
important because  the  ED data 
(due to finite-size effects)  do not fix the position of the  tip.  

\begin{figure}[hbt]
\samepage
\begin{center}
\includegraphics[width=8cm]{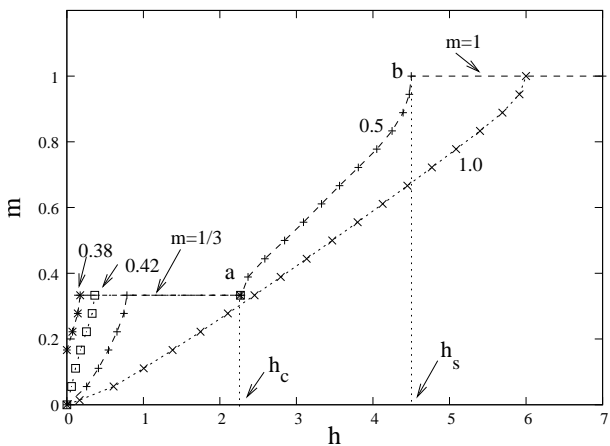}
\caption{\label{fig2}
Magnetization curves of the model (\ref{h}) for
different values of the frustration parameter 
$J_2$ ($=0.38,0.42,0.5$, and $1$).  $m=M/M_s$ is the reduced
magnetization,
$M$ being the magnetic moment per rung, and $M_s=3/2$. 
The symbols mark the midpoints
of the magnetization steps, as obtained from the ED data for $L=12$.
For clarity, we do  not show the data points for $J_2=0.38$ and $0.42$
in the interval $h_{c}<h<h_{s}$ since they  closely follow 
the presented data for $J_2=0.5$.
}
\end{center}
\end{figure}

To compare with the phase diagram of the boson 
Hubbard model, recall that  the magnetic field 
$h$ plays the role of a chemical potential  and the
magnetization $M$ is an analog of the conserved 
density of  particles  $\rho$. As is well known,
the  Luttinger liquid  phase is an analog of the 
superfluid phase in the boson model, 
and the plateau phases can be viewed  as Mott-insulator  
phases,\cite{totsuka}  characterized by an integer $\rho$, 
a finite gap for particle-hole excitations,
and  zero compressibility, $\kappa=\partial
\rho/\partial\mu$=0.  
Examples of magnetization curves $m(h)$ for different parameters 
$J_2$ are presented in Fig. \ref{fig2}. These are analogs
of the $\rho (\mu)$ curves for the 1D boson Hubbard model.\cite{bartouni}
Note that in  the  case  $J_2=1$ the curve  $m(h)$ describes
the reduced magnetization of the spin-$\frac{3}{2}$ antiferromagnetic Heisenberg
chain.\cite{note} The magnetization curves also indicate a  divergence 
of the zero-field magnetic susceptibility $\chi=\chi(J_2)$ 
as $J_2$ approaches the isolated critical point $J_c$ ($J_2\to J_c\pm 0$), 
in accordance with  the generic Luttinger liquid relation\cite{konik} 
$\chi(h,J_2)=K(h,J_2)/[\pi v_s(h,J_2)]$  and the 
observation that  the spin-wave velocity  
$v_s(0,J_2)\rightarrow 0$ as $J_2 \rightarrow J_c \pm0$.\cite{ivanov1} 
Here $K(h,J_2)$ is the Luttinger liquid parameter at finite $h$ and $J_2$.
Up to now, only a few studies analyzing  such
ferromagnetic quantum critical points
in 1D have been published.\cite{daul,yang,sengupta}
In particular, as  recently pointed out by Sengupta and Kim,\cite{sengupta}, 
in the isotropic case the ferromagnetic transition at $J_c$ can be
different from the ferromagnetic transition in the Ising limit\cite{yang}
due to quartic terms in the effective action which couple the longitudinal 
mode to the transverse spin-wave modes. 
The magnetization curves also indicate
that  the canted magnetic phase is characterized by 
a finite susceptibility $\chi(0,J_2)$, as can  be expected since
this state may be viewed as a kind of ferromagnetic Luttinger liquid
phase.\cite{bartosch}

Discussed instabilities at $h_c$ and $h_s$ share the same physics as
those at  the  lower ($h_{c1}$) and upper ($h_{c2}$) 
critical fields in Haldane-gap  chains  in a magnetic field.\cite{maisinger} 
In particular, the  instability  at $h_c$  may be viewed  
as a  condensation of a  dilute gas of magnon excitations,  
as in the case of Haldane-gap  chains  slightly above the lower 
critical  field $h_{c1}$.\cite{takahashi,affleck,sorensen} 
At the  critical field  $h_c$,  
the gap of the AFM mode  $\omega^{(-)}_{0}(k)$  
vanishes and, as a result, a macroscopic  amount of AFM 
magnons condense onto the ground state. The collapse of 
quasiparticles is prevented by the repulsive on-site 
quasiparticle interaction. For  $h>h_{c}$,  
the magnetization $M$ continuously 
exceeds the quantized value $1/2$. 
The instabilities related to the FM modes, as in the case
of a Haldane  chain at the upper critical field 
$h_{c2}$, are  expected to share the same critical 
properties.\cite{konik,essler} In  fact,
the transitions discussed  are of a  commensurate-incommensurate 
type,\cite{pokrovsky} like  the insulator-superfluid transition 
in the 1D boson Hubbard model on the phase boundaries where the density
continuously changes from $\rho=1$ to some incommensurate value
$\rho > 1$ or $\rho < 1$. Since magnons behave as a dilute gas of
hard-core bosons, in a first  approximation the specific 
form of the interaction is irrelevant.\cite{sorensen,sachdev}  
This results, in particular, in the  universal behavior 
of the Haldane chain 
magnetization\cite{tsvelik,affleck}
$M=\left[ 2\Delta_H (h-h_{c1})\right]^{1/2}/(\pi v_s)$, 
where $\Delta_H$ is the Haldane gap and $v_s$ ($ \approx 2.49$)
is the spin-wave  velocity. Similar universal expressions 
can be obtained close to the phase boundaries of the
$(J_2,h)$ diagram of our model.
Using the equivalence of 1D hard-core bosons and  free spinless fermions,
this implies, in particular,  the  Luttinger liquid 
parameter $K=1$  on the  entire boundary 
of the $M=1/2$ phase as well as on the $h_s(J_2)$
boundary, excluding the special point at
the tip of   the region occupied by the
$M=1/2$ phase. Respectively, the  spin-spin correlation
exponents take the values $\eta_z=2$ and $\eta_x=\frac{1}{2}$,
and the magnetization exponent $\beta=1/2$.
Based on the established  picture of the low-lying
excitations  close to the point $(J_2^*,h^*)$  
and using the  analogy with the  boson  phase diagram, 
it is plausible to assume  that the latter critical point
describes a Berezinskii-Kosterlitz-Thouless (BKT) 
type transition. 
Although our  ED data do not fix the position of the tip, 
it is easy to indicate the typical pointed-type shape of the 
boundary\cite{kuhner} resulting from the specific 
BKT particle-hole excitation  gap $\Delta_{ph}$ close to 
this transition: 
$\Delta_{ph}\sim \exp \left(  const/\sqrt{J_2^*-J_2}\right)$.
As might  be expected, the SWT does not reproduce the correct
behavior of the gap $\Delta_{ph}$ close to this special point.

In conclusion, we have demonstrated that a  
generic mixed-spin ladder   exhibits
a   magnetic phase diagram which is closely
related  to the one  of the boson Hubbard model
in one spatial dimension. Similar magnetic phase 
diagrams may be expected for  arbitrary 
quantum spin numbers $s_1$ and $s_2$, provided that 
$s_1\neq s_2$ and $s_1+s_2=$half-integer.

\begin{acknowledgments}
This work was  partially supported by the Bulgarian
Science Foundation, Grant No. 1414/2004.   
\end{acknowledgments}


\begin{thebibliography}{}
\bibitem{note}
As shown in Ref. \onlinecite{ivanov1}, at $J_2=1$ the ladder model
(\ref{h}) is equivalent in the low-energy sector to the 
spin-$\frac{3}{2}$ antiferromagnetic Heisenberg chain.
\end{thebibliography}
\end{document}